\documentclass[twocolumn,aps,prl,superscriptaddress,amsfonts]{revtex4-1}

\usepackage{amsmath}
\usepackage{graphicx}
\usepackage{bm}
\usepackage{array}
\usepackage{dcolumn}
\usepackage{hepparticles}
\usepackage{heppennames}
\usepackage{hepnicenames}
\usepackage{epstopdf}

\begin{document}

\newcommand{\ket}[1]{{| {#1} \rangle}}

\newcommand{\FigureOne}{
\begin{figure}[htbp!]
\includegraphics*[width=\columnwidth]{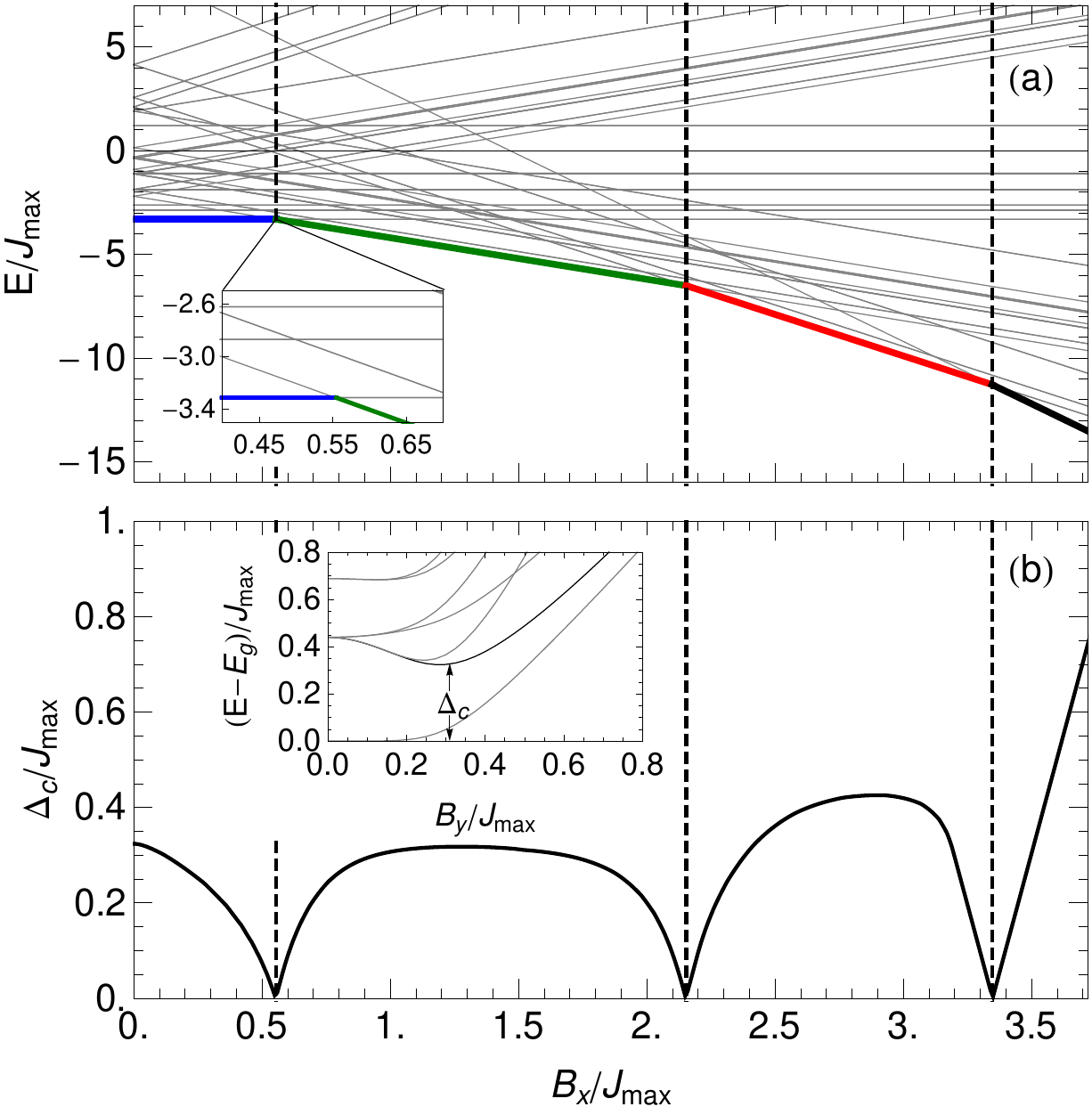}
\caption{(a) Low-lying energy eigenvalues of Eqn. \ref{eqn:Hamiltonian} for $B_y=0$ and $N=6$, with the long-range $J_{i,j}$ couplings determined from experimental conditions (see text). Level crossings (inset) indicate the presence of first-order phase transitions in the ground state. (b) The critical gap $\Delta_c$ shrinks to zero at the three phase transitions (vertical dashed lines). Inset: low-lying energy levels of Eqn. $\ref{eqn:Hamiltonian}$ with $B_x=0$.}
\label{fig:EnergyLevels}
\end{figure}
}

\newcommand{\FigureTwo}{
\begin{figure}[t!]
\includegraphics*[width=\columnwidth]{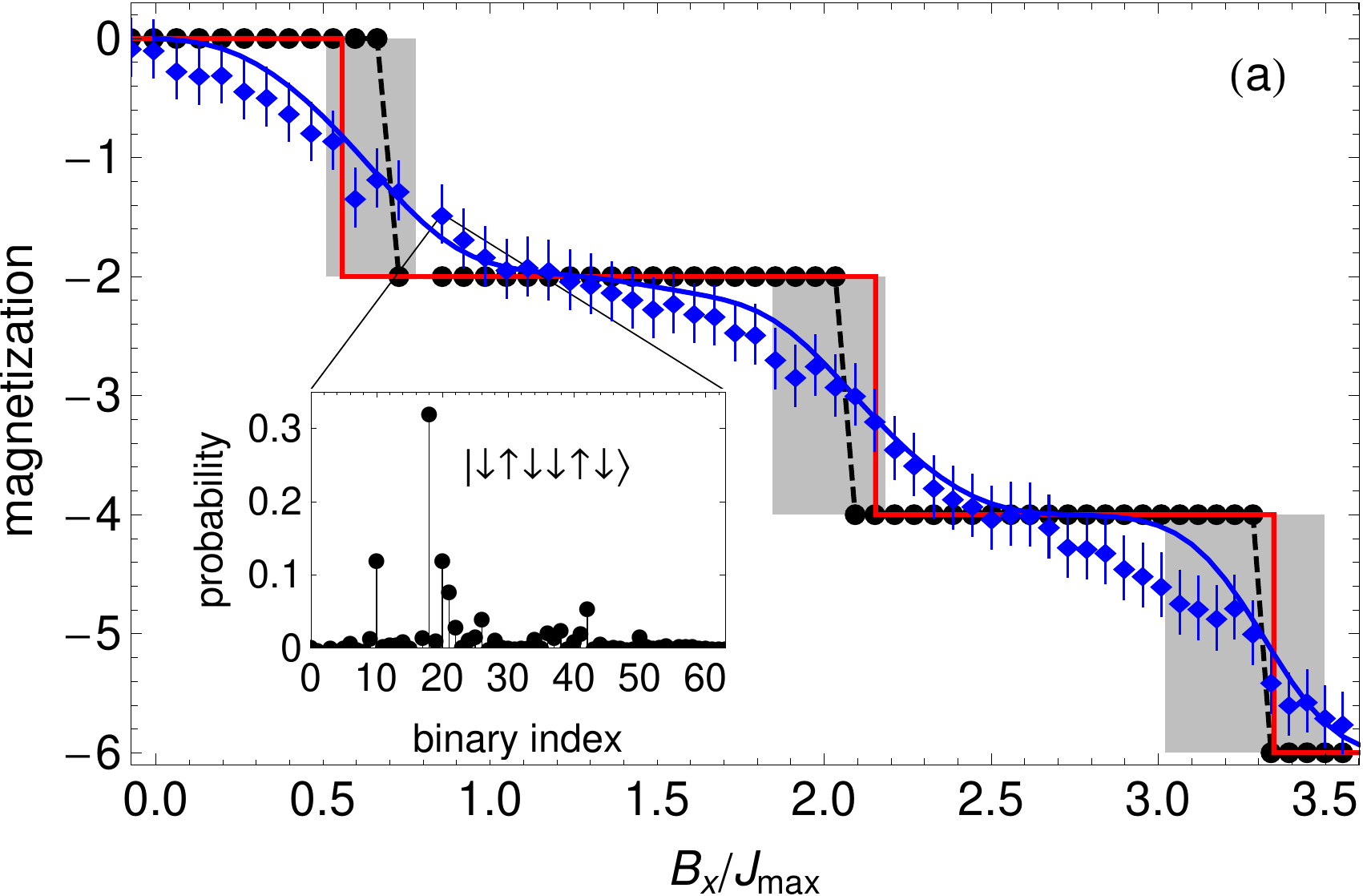}
\includegraphics*[width=\columnwidth]{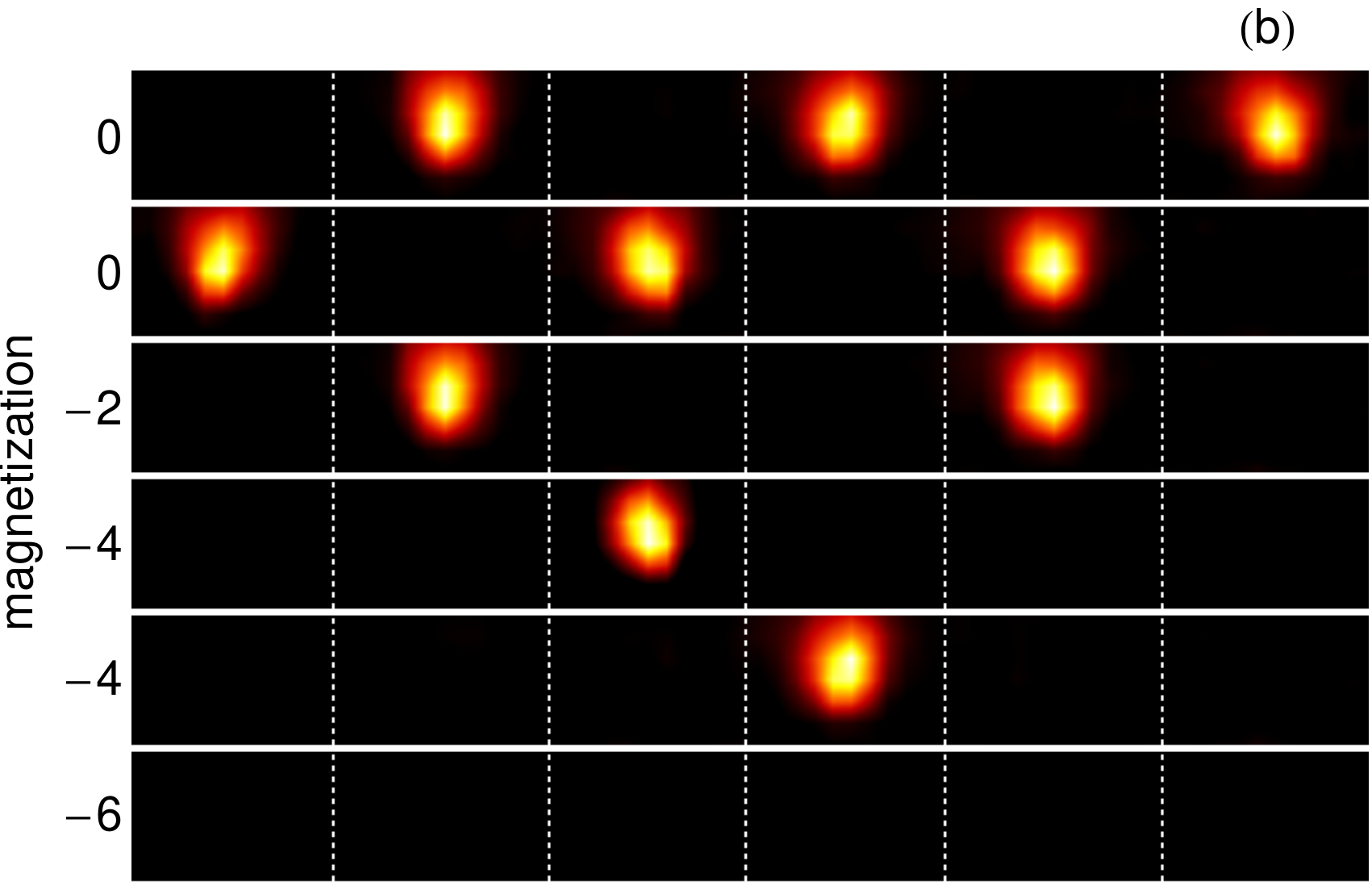}
\caption{(a) Magnetization ($m_x=N_{\uparrow}-N_{\downarrow}$) of 6 ions for increasing axial field strength. Red, solid: magnetization of the calculated ground state, with the step locations indicating the first-order phase transitions. Blue diamonds: average magnetization of 4000 experiments for various $B_x$. Blue, solid: magnetization calculated by numerical simulation using experimental parameters. Black, dashed: magnetization of the most probable state (see inset) found at each $B_x$ value. Gray bands indicate the experimental uncertainty in $B_x/J_{\text{max}}$ at each observed phase transition. (b) Linearly interpolated camera images of the ground states found at each step in (a): $\ket{\downarrow\uparrow\downarrow\uparrow\downarrow\uparrow}$ and $\ket{\uparrow\downarrow\uparrow\downarrow\uparrow\downarrow}$ ($m_x=0$), $\ket{\downarrow\uparrow\downarrow\downarrow\uparrow\downarrow}$ ($m_x=-2$), $\ket{\downarrow\downarrow\uparrow\downarrow\downarrow\downarrow}$ and $\ket{\downarrow\downarrow\downarrow\uparrow\downarrow\downarrow}$ ($m_x=-4$), and $\ket{\downarrow\downarrow\downarrow\downarrow\downarrow\downarrow}$ ($m_x=-6$).}
\label{fig:StepsSixIon}
\end{figure}
}

\newcommand{\FigureThree}{
\begin{figure}[t]
\includegraphics*[width=\columnwidth]{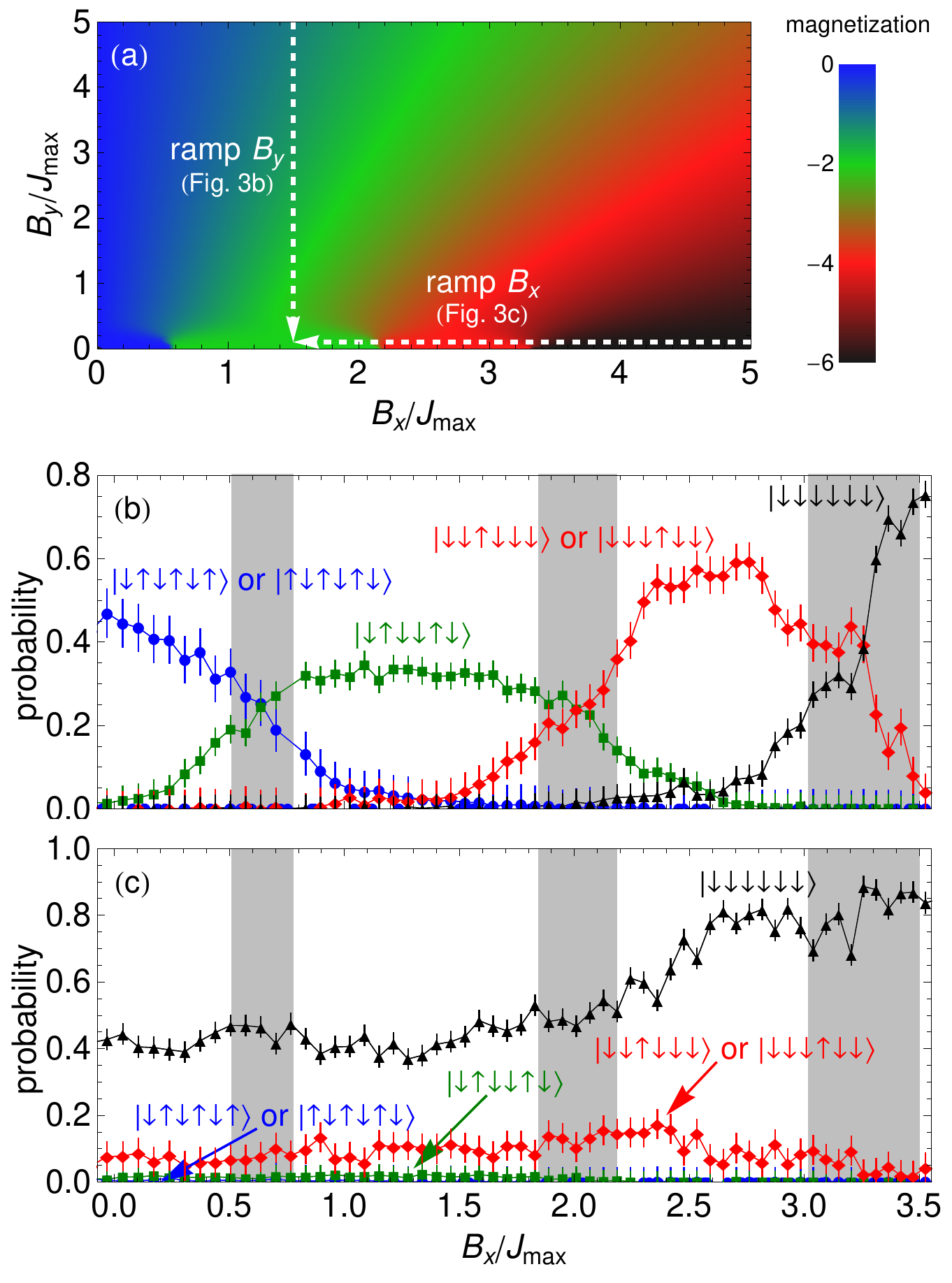}
\caption{(a) Ground state phase diagram of the system, along with two different trajectories that end at the same value of $B_x$. (b) Probabilities of the 4 different ground state spin phases when $B_y$ is ramped in a 6-ion system. Blue dots: $\ket{\downarrow\uparrow\downarrow\uparrow\downarrow\uparrow}$ or $\ket{\uparrow\downarrow\uparrow\downarrow\uparrow\downarrow}$. Green squares: $\ket{\downarrow\uparrow\downarrow\downarrow\uparrow\downarrow}$. Red diamonds: $\ket{\downarrow\downarrow\uparrow\downarrow\downarrow\downarrow}$ or $\ket{\downarrow\downarrow\downarrow\uparrow\downarrow\downarrow}$. Black triangles: $\ket{\downarrow\downarrow\downarrow\downarrow\downarrow\downarrow}$. Gray bands are the experimental uncertainties of the phase transition locations. (c) Probabilities of creating the 4 different ground states when $B_x$ is ramped. Most of the ground states are classically inaccessible in our zero temperature system.}
\label{fig:stateprobs}
\end{figure}
}

\newcommand{\FigureFour}{
\begin{figure}[t!]
\includegraphics*[width=\columnwidth]{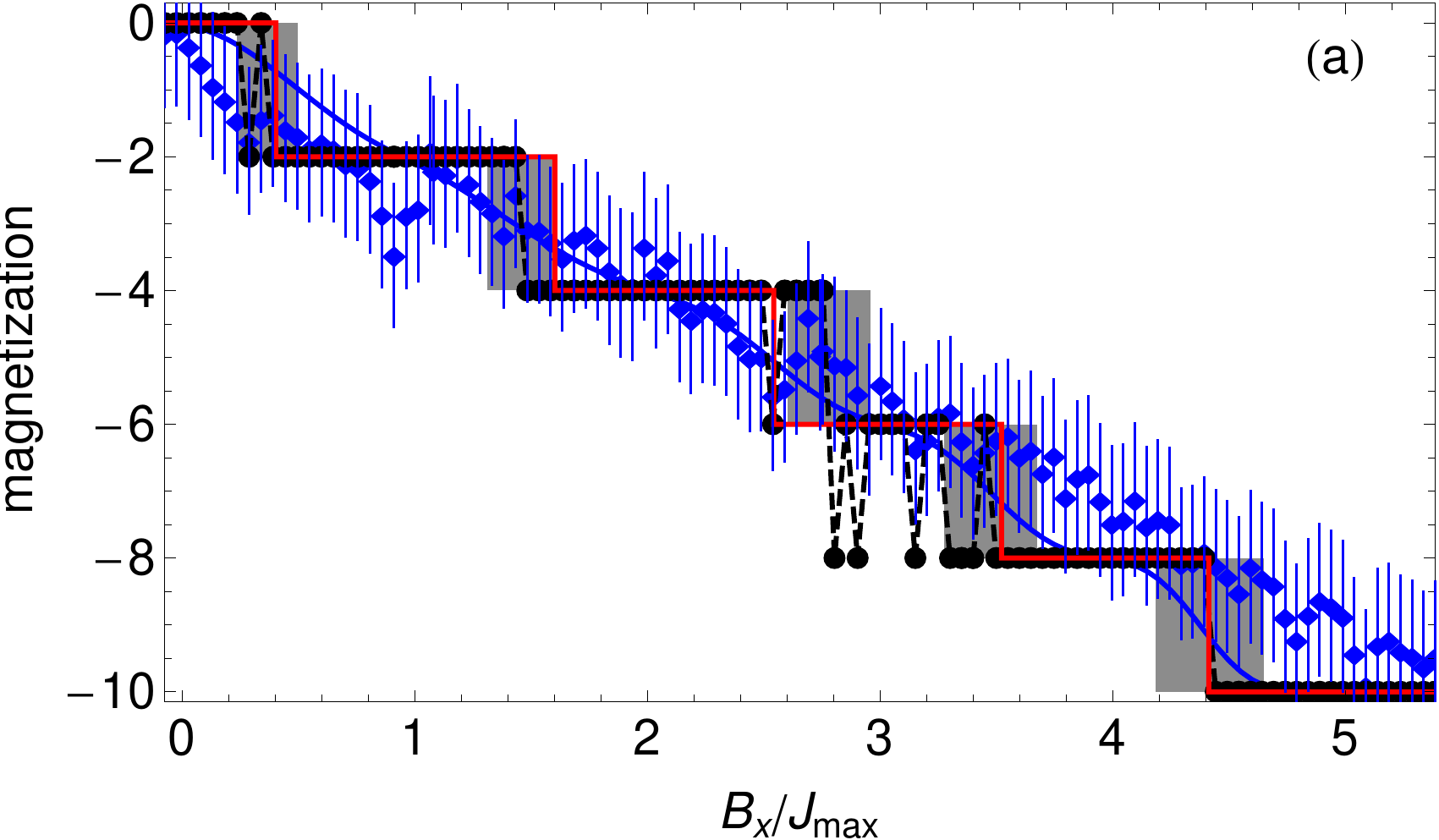}
\includegraphics*[width=\columnwidth]{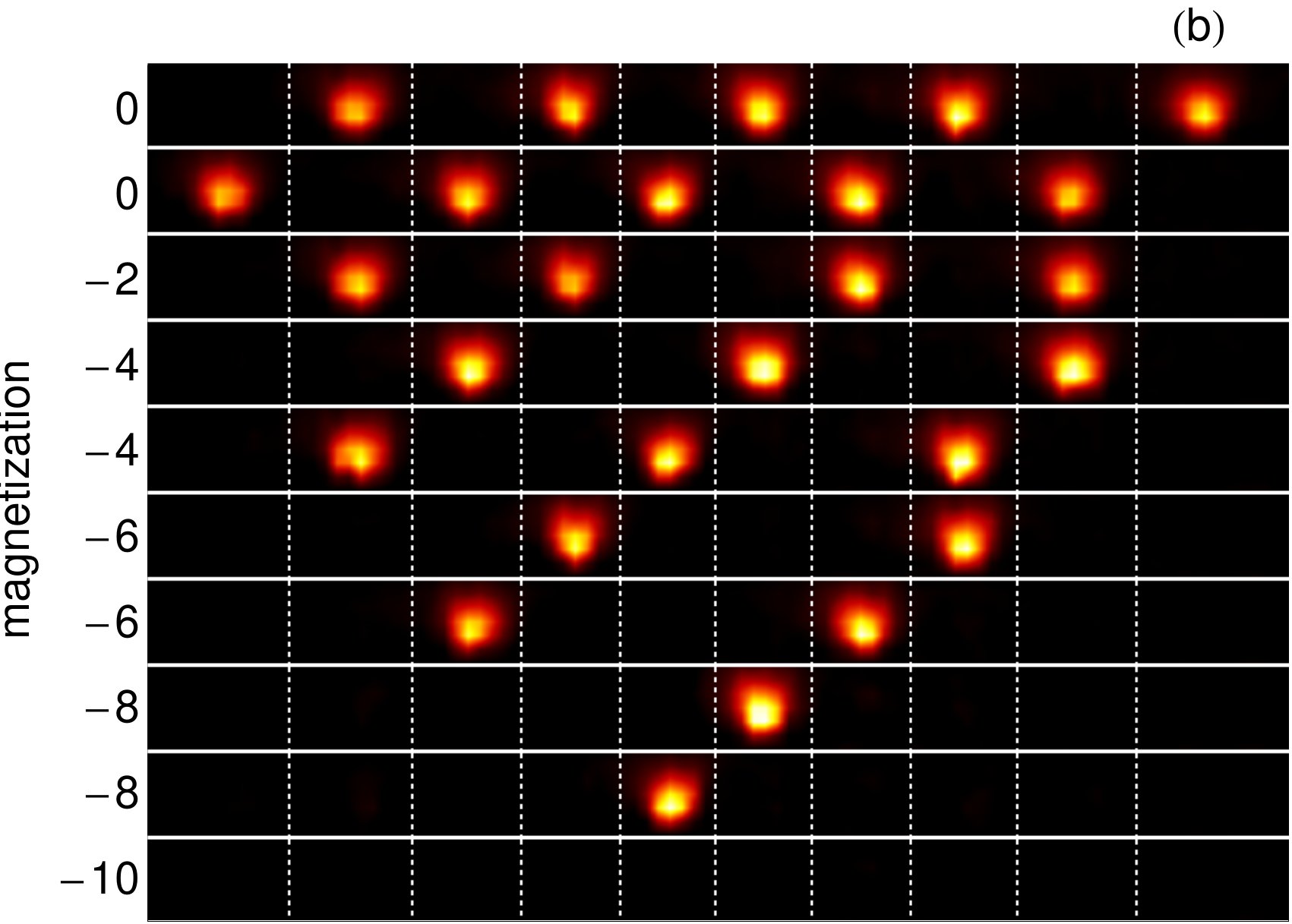}
\caption{(a) Magnetization of a chain of 10 ions for increasing axial field strength. The red, blue, and black curves correspond to the theoretical magnetization, average measured and simulated magnetization, and magnetization of the most probable state (respectively) for increasing $B_x$. Gray bands show the measurement uncertainty of the phase transition locations. (b) Linearly interpolated camera images of the ground state spin configuration at each magnetization.}
\label{fig:StepsTenIon}
\end{figure}
}

\title{Quantum Catalysis of Magnetic Phase Transitions in a Quantum Simulator}

\author{P. Richerme$^1$, C. Senko$^1$, S. Korenblit$^1$, J. Smith$^1$, A. Lee}
\affiliation{Joint Quantum Institute, University of Maryland Department of Physics and National Institute of Standards and Technology, College Park, MD  20742}

\author{R. Islam}
\affiliation{Department of Physics, Harvard University, Cambridge, MA 02138}

\author{W. C. Campbell}
\affiliation{Department of Physics and Astronomy, University of California, Los Angeles, CA  90095}

\author{C. Monroe}
\affiliation{Joint Quantum Institute, University of Maryland Department of Physics and National Institute of Standards and Technology, College Park, MD  20742}

\date{\today}

\begin{abstract}
We control quantum fluctuations to create the ground state magnetic phases of a classical Ising model with a tunable longitudinal magnetic field using a system of 6 to 10 atomic ion spins. 
Due to the long-range Ising interactions, the various ground state spin configurations are separated by multiple first-order phase transitions, which in our zero temperature system cannot be driven by thermal fluctuations. 
We instead use a transverse magnetic field as a quantum catalyst to observe the first steps of the
complete fractal devil's staircase, which emerges in the thermodynamic limit and can be mapped to a large number of many-body and energy-optimization problems.

\end{abstract}

\maketitle
Quantum simulators, in which a well-controlled quantum system is used to simulate a system of interest \cite{FeynmanQSIM,LloydQSIM}, offer the promise of calculating ground state or dynamical properties of a Hamiltonian that would otherwise prove classically intractable. For highly frustrated and long-ranged spin systems \cite{BinderSpinGlasses,BalentsSpinLiquids,BramwellSpinIce,DawsonFrustration}, or for studies of non-equilibrium evolution, numerical calculations are in many cases limited to only a few dozen sites \cite{Lanczos,SandvikScaling,MassivelyParallelQCSim}, motivating simple quantum simulations in a variety of contexts \cite{BlochSingleSpinAddressing,EsslingerProbingUltracoldFermions, GreinerAFM,NMRQuantumSimulation,PhotonQuantumSimulatorReview,PhotonicBosonSampling, QSIMNature2010,BlattDigitalQSIM,Britton2012}. Quantum simulations of lattice spin models using trapped atomic ions have shown particularly encouraging results, with the number of interacting spins increasing from 2 to 16 \cite{FriedenauerQSIM,QSIM2013Science} in the past 5 years.

Quantum simulators require exquisite isolation from their environment to realize long coherence times and high-fidelity readout and control. Such isolation allows the system to be initialized into a pure state and remain decoupled from a thermal bath over the duration of the simulation. Many previous quantum simulations have investigated the transverse-field Ising model \cite{QSIMNature2010,QSIMPRB2010,QSIMNatureComm2011,QSIM2013Science}, pairing an effectively zero-temperature system with engineered spin-spin couplings and magnetic fields. In these experiments, the paramagnetic and ordered spin phases are separated by a quantum phase transition driven by quantum fluctuations from the transverse field.

In contrast, we consider here a classical spin model: the longitudinal-field Ising model with convex, long-range antiferromagnetic (AFM) interactions. For increasing longitudinal field strength $B_x$, the system exhibits many distinct ground state phases separated by first-order classical phase transitions. Yet even for just a few spins, the various ground states at different $B_x$ are classically inaccessible in a physical system at or near zero temperature due to the absence of thermal fluctuations to drive the phase transitions \cite{SachdevBook}. 

In this Letter, we show how to create these classically inaccessible ground states by applying a transverse field (which does not commute with the longitudinal-field Ising Hamiltonian) to introduce quantum fluctuations. Using $N=6$ or $N=10$ spins, we use this technique to experimentally identify the locations of the multiple classical phase transitions and to preferentially populate each of the classical ground states that arise for varying strengths of the longitudinal field. We observe a ground state spin ordering that reveals a Wigner-crystal spin structure \cite{Wigner1934}, maps on to a number of energy minimization problems \cite{QUBO,DWaveProteinFolding}, and shows the first steps of the complete devil's staircase \cite{PerBakDevilsStaircasePRL} which would emerge in the $N\rightarrow\infty$ limit.

The system is described by the Hamiltonian
\begin{equation}
H=\sum_{i<j} J_{i,j} \sigma_x^{(i)} \sigma_x^{(j)} + B_x \sum_{i} \sigma_x^{(i)} + B_y(t) \sum_{i} \sigma_y^{(i)}
\label{eqn:Hamiltonian}
\end{equation}
where $J_{i,j}>0$ gives the strength of the AFM Ising coupling between spins $i$ and $j$, $B_x$ is the longitudinal magnetic field, $B_y(t)$ is a time-dependent transverse field, and $\sigma_{\alpha}^{(i)}$ is the Pauli spin operator for the $i$th particle along the $\alpha$ direction. The couplings $J_{i,j}$ and field magnitudes $B_x$ and $B_y(t)$ are given in units of angular frequency, with $\hbar=1$. At \mbox{$t=0$} the spins are initialized along the total magnetic field $\vec{B}=B_x \hat{x}+B_y(0) \hat{y}$, with $B_y(0) \gg J$, which is the paramagnetic, instantaneous ground state of the Hamiltonian (Eqn. \ref{eqn:Hamiltonian}) to good approximation. The transverse field $B_y(t)$ is then slowly ramped to 0, crossing a quantum phase transition driven by quantum fluctuations, to preferentially create the ground state of the classical longitudinal-field Ising Hamiltonian.

\FigureOne

Fig. \ref{fig:EnergyLevels}(a) shows the energy eigenvalues of the Hamiltonian (Eqn. \ref{eqn:Hamiltonian}) with $B_y=0$ for a system of 6 spins. The ground state passes through three level crossings as $B_x$ is increased from 0, indicating three classical first-order phase transitions separating four distinct spin phases. For each $B_x$, there is a quantum critical point at some finite $B_y$ characterized by a critical gap $\Delta_c$ (inset of Fig. \ref{fig:EnergyLevels}(b)). When $B_x$ is near a classical phase transition, the near energy-degeneracy of spin orderings shrinks the quantum critical gap, as shown in Fig. \ref{fig:EnergyLevels}(b).

Long-range interactions give rise to many more ground state spin phases than does a nearest-neighbor-only Ising model. Consider a nearest-neighbor AFM model with $N$ total spins and a ground-state ordering $\ket{..\downarrow\uparrow\downarrow\uparrow\downarrow\uparrow..}$. An excited state at $B_x=0$ may have an additional spin polarized along $\ket{\downarrow}$, either by making a kink of type $\ket{..\downarrow\uparrow\downarrow\downarrow\uparrow\downarrow..}$ or a spin defect of type $\ket{..\downarrow\uparrow\downarrow\downarrow\downarrow\uparrow..}$. The interaction energy gain of making $n$ kinks is $2nJ$, while the field energy loss is $2nB_x$. At $B_x/J=1$, multiple energy levels intersect to give a first-order phase transition. Similarly, the energy gain of making $n$ spin defects is $4nJ$ and the loss is $2nB_x$, so a second phase transition occurs at $B_x/J=2$. Only three different ground state spin phases are observable as $B_x$ is varied from $0 \rightarrow \infty$, independent of $N$, and there is a large degeneracy of spin eigenstates at the phase transitions. The presence of long-range interactions lifts this degeneracy and admits $[N/2]+1$ distinct spin phases with $\{0,1,\ldots,[N/2]\}$ spins in state $\ket{\uparrow}$, where $[N/2]$ is the integer part of $N/2$.

The effective spin system is encoded in a linear chain of trapped $^{171}$Yb$^+$ ions with zero effective spin temperature. The spin states $\ket{\uparrow}_z$ and $\ket{\downarrow}_z$ are represented by the hyperfine clock states $^2S_{1/2}$ $\ket{F=1,m_F=0}$ and $\ket{F=0,m_F=0}$, respectively, which have a frequency splitting of \mbox{$\omega_{S}/2\pi=12.642819$ GHz} \cite{YbDetection}. A weak magnetic field of $\sim5$ G defines the quantization axis. The states are detected by illuminating the ions with laser light resonant with the $^2S_{1/2}$ to $^2P_{1/2}$ cycling transition at 369.5 nm and imaging the spin-dependent fluorescence. Either $N=6$ or $N=10$ ions are confined in a three-layer rf Paul trap with a center-of-mass axial trap frequency $f_{Z}=0.7$ MHz and transverse frequencies $f_{X}=4.8$ MHz and $f_{Y}=4.6$ MHz and interact with each other via their collective modes of motion. 

The Ising couplings $J_{i,j}$ are generated by globally irradiating the ions with two off-resonant $\lambda\approx 355$ nm laser beams which drive stimulated Raman transitions \cite{Hayes2010,Campbell2010}. The beams intersect at right angles so that their wavevector difference $\Delta \vec{k}$ points along the $X$-direction of transverse ion motion, perpendicular to the linear chain. Acousto-optic modulators imprint beatnote frequencies of $\omega_{S}\pm\mu$ between the beams, imparting a spin-dependent optical dipole force at frequency $\mu$ \cite{PorrasCiracQSIM,MolmerSorensen}. In the limit where the beatnotes are sufficiently far from the transverse normal modes $\omega_m$, we obtain the spin-spin coupling
\begin{equation}
\label{eqn:Jij}
J_{i,j}=\Omega_i \Omega_j \frac{\hbar (\Delta\vec{k})^2}{2M}\sum_m \frac{b_{i,m} b_{j,m}}{\mu^2-\omega_m^2}
\end{equation}
in the Lamb-Dicke limit, where $\Omega_i$ is the Rabi frequency of the $i$th ion, $M$ is the single ion mass, and $b_{i,m}$ is the normal-mode transformation matrix for ion $i$ in mode $m$ \cite{QSIMPRL2009}. The Ising interactions are long-range and fall off approximately as $J_{i,j}\sim J_{\text{max}}/|i-j|^{\alpha}$, where $J_{\text{max}}$ is typically $2\pi\times$0.6-0.7 kHz, $\alpha=0.94$ for $N=6$, and $\alpha=0.83$ for $N=10$.

The effective transverse and longitudinal magnetic fields $B_y(t)$ and $B_x$ drive Rabi oscillations between the spin states $\ket{\downarrow}_z$ and $\ket{\uparrow}_z$. Each effective field is generated by a pair of Raman laser beams with a beatnote frequency of $\omega_{S}$, with the field amplitude determined by the beam intensities. The field directions are controlled through the beam phases relative to the average phase $\varphi$ of the two sidebands which give rise to the $\sigma_x\sigma_x$ interaction in Eqn. \ref{eqn:Hamiltonian}. In particular, an effective field phase offset of $0^\circ$ ($90^\circ$) relative to $\varphi$ generates a $\sigma_y$ ($\sigma_x$) interaction.

Each experiment begins with 3 ms of Doppler cooling, followed by optical pumping to the state $\ket{\downarrow\downarrow\downarrow..}_z$ and $100~\mu$s of Raman sideband cooling that prepares the motion of all modes along $\Delta \vec{k}$ in the Lamb-Dicke limit. The spins are then coherently rotated into the equatorial plane of the Bloch sphere so that they point along $\vec{B}=B_x \hat{x}+B_y(0) \hat{y}$, with $B_x$ varied between different simulations. The Hamiltonian (Eqn. \ref{eqn:Hamiltonian}) is then switched on at $t=0$ with the chosen value of $B_x$ and $B_y(0)=5 J_{\text{max}}$. The transverse field (which provides the quantum fluctuations) is ramped down to  $B_y\approx0$ exponentially with a time constant of $600~\mu$s and a total time of 3 ms, which sacrifices adiabaticity in order to avoid decoherence effects. At $t=3$ ms, the Hamiltonian is switched off and the $x-$component of each spin is measured by applying a global $\pi/2$ rotation about the $\hat{y}$ axis, illuminating the ions with resonant light, and imaging the spin-dependent fluorescence using an intensified CCD camera. Experiments are repeated 4000 times to determine the probability of each possible spin configuration. We compensate for known detection errors ($\epsilon=7\%$ for a single spin) by multiplying a matrix describing the expected multi-spin error by the vector containing the measured probability of each spin configuration \cite{CorrectingDetectionErrors, QSIM2013Science}.

\FigureTwo

We investigate the order parameter of net magnetization along $x$, $m_x=N_{\uparrow}-N_{\downarrow}$, as we tune the longitudinal field strength. The magnetization of the ground state spin ordering of Eq. \ref{eqn:Hamiltonian} is expected to yield a staircase with sharp steps at the phase transitions (red line in Fig. \ref{fig:StepsSixIon}(a)) when $B_y=0$ \cite{PerBakDevilsStaircasePRL}. The experimental data (blue points in Fig. \ref{fig:StepsSixIon}(a)) show an averaged magnetization with heavily broadened steps due largely to the non-adiabatic exponential ramp of the transverse field. The deviation from sharp staircase-like behavior is predicted by numerical simulations (solid blue line in Fig. \ref{fig:StepsSixIon}(a)) which use our experimental parameters and ramp profiles. Differences between theory and experiment are largest near the phase transitions, where excitations are easier to make due to the shrinking quantum critical gap (Fig. \ref{fig:EnergyLevels}(b)).

Nevertheless, we extract the ground state spin configuration at each value of $B_x$ by looking at the probability distribution of all spin states and selecting the most prevalent state (inset of Fig. \ref{fig:StepsSixIon}(a)) \cite{QSIMOptimalRamp}. The magnetization of the spin states found by this method (black points in Fig. \ref{fig:StepsSixIon}(a)) recover the predicted staircase structure. The steps in the experimental curve agree with the calculated phase transition locations to within experimental error (gray bands in Fig. \ref{fig:StepsSixIon}(a)), which accounts for statistical uncertainty due to quantum projection noise and estimated drifts in the strengths of $J_{i,j}$, $B_x$, and $B_y$.

Fig. \ref{fig:StepsSixIon}(b) shows approximately 1000 averaged camera images of the most probable spin configuration observed at each plateau in Fig. \ref{fig:StepsSixIon}(a). Each box contains an ion that scatters many photons when in the state $\ket{\uparrow}$ and essentially no photons when in the state $\ket{\downarrow}$. The observed spin orderings in Fig. \ref{fig:StepsSixIon}(b) match the calculated ground states at each magnetization, validating the technique of using quantum fluctuations to preferentially create these classically inaccessible ground states. (For magnetizations of $0$ and $-4$, two ground state orderings are observed due to the left-right symmetry of the spin-spin interactions.)

\FigureThree

To further illustrate the necessity of using quantum fluctuations to catalyze the magnetic phase transitions, we consider alternate ramp trajectories for reaching a final chosen value of $B_x$. Fig. \ref{fig:stateprobs}(a) shows the ground state phase diagram of the Hamiltonian (Eqn. \ref{eqn:Hamiltonian}), with the sharp classical phase transitions visible along the bottom axis ($B_y/J_{\text{max}}=0$). In addition, it shows two possible trajectories through the phase diagram that start in a paramagnetic ground state (which is easy to prepare experimentally) and end at the same value of $B_x$ with $B_y=0$.

We have already used the first trajectory, in which $B_x$ is fixed and $B_y$ is ramped from $5J_{\text{max}}$ to 0, to experimentally verify the locations of the 3 classical phase transitions and experimentally create the 4 different ground state phases. In Fig. \ref{fig:stateprobs}(b), we plot the probability of creating each ground state as a function of $B_x$ and find populations of $\sim40-80\%$. We observe a smooth crossover between the four ground state phases, with the classical phase transitions occuring at the crossing points. This arises since distinct spin eigenstates have degenerate energies at the phase transition, causing the quantum critical gap between them to close and allowing quantum fluctuations to populate both states equally (see Fig. \ref{fig:EnergyLevels}).

The second trajectory in Fig. \ref{fig:stateprobs}(a) is purely classical, with $B_y$ set to 0. The spins are initialized into the state $\ket{\downarrow\downarrow\downarrow\downarrow\downarrow\downarrow}$, and $B_x$ is ramped from $5J_{\text{max}}$ to its final value at a rate of $5J_{\text{max}}/3$ ms. Fig. \ref{fig:stateprobs}(c) shows that in a classical system without thermal or quantum fluctuations, the phase transitions remain undriven and the initial state $\ket{\downarrow\downarrow\downarrow\downarrow\downarrow\downarrow}$ remains dominant for all values of $B_x$. The ground state phases with magnetization 0 and $-2$ (blue and green in Fig. \ref{fig:stateprobs}(c)) are separated from the initial state by several classical phase transitions and have essentially zero probability of being created.

\FigureFour

The technique of introducing quantum fluctuations may be applied to larger chains of ions and is demonstrated in Fig. \ref{fig:StepsTenIon}(a) with $N=10$. As before, we can post-select the most prevalent of the $2^{10}=1024$ possible states and plot their magnetization, revealing $N/2+1=6$ different plateaus. However, the critical gap $\Delta_c$ is much smaller for a 10-ion system and ramping $B_y(t)$ induces many more excitations. Unlike the $\sim40-80\%$ ground state population for the 6-ion case, the 10-ion ground states are made with a probability of only $\sim5\%$, which approaches the level of statistical and experimental error in the simulation. Even so, we can still experimentally determine the classical phase transition locations and find good agreement with theory. Longer coherence times (which would enable slower transverse-field ramps) or optimized ramp profiles \cite{QSIMOptimalRamp} would likely improve the ground state fraction in larger systems.

For the 10-ion chain, Fig. \ref{fig:StepsTenIon}(b) shows the interesting ground state spin structure that emerges as $B_x$ is varied. For a given $B_x$ and associated number of bright ions $q$, the ground state spin configuration of Eqn. \ref{eqn:Hamiltonian} (with $B_y=0$) solves the minimization problem of finding the lowest energy arrangement of $q$ charged particles on $N$ lattice sites. The creation of such periodic spin structures realizes a generalized Wigner crystal \cite{HubbardTCNQ}, mapping the configuration of a cold, low-density electron gas onto our zero temperature spin system. As the system size $N\rightarrow\infty$, the staircase structure in magnetization becomes a fractal that arises since every rational filling factor (of which there are infinitely many) is the ground state for some value of $B_x$ \cite{PerBakDevilsStaircasePRL}.

In conclusion, we have demonstrated that when there are no thermal fluctuations to drive classical phase transitions, quantum fluctuations may be introduced to create otherwise inaccessible phases. This provides a tool for making ground states which solve energy optimzation problems or studying phase transitions in classical systems at or near zero temperature. The technique should equally well apply to other classical systems whenever the absence of thermal fluctuations prevents the system from finding its ground state.

The authors wish to thank Robijn Bruinsma for helpful discussions. This work is supported by the U.S. Army Research Office (ARO) Award W911NF0710576 with funds from the DARPA Optical Lattice Emulator Program, ARO award W911NF0410234 with funds from the IARPA MQCO Program, and the NSF Physics Frontier Center at JQI.

\bibliographystyle{prsty}
\bibliography{qsimrefs}
\end{document}